\documentclass{aa}
\usepackage{graphicx}
\begin{document}

   \title{Letter to the Editor:\\
    Wind anisotropies and GRB progenitors}

%\subtitle{GRB progenitors} 
\titlerunning{Wind anisotropies and GRB progenitors}

\author{Georges Meynet \& Andr\'e Maeder}

     \institute{Geneva Observatory, University of Geneva, CH--1290 Sauverny, Switzerland\\
              email: georges.meynet@obs.unige.ch\\
             email: andre.maeder@obs.unige.ch
               }

   \date{Received  / Accepted }

   \offprints{georges.meynet@obs.unige.ch}

\abstract{}
{We study the effect of wind anisotropies on the stellar evolution leading to collapsars.} 
{Rotating  models of a 60 M$_\odot$ star with $\Omega/\Omega_{\rm crit}=0.75$ on the ZAMS, accounting for shellular rotation and a magnetic field, with and without wind anisotropies, are computed at $Z$=0.002 until the end of the core He-burning phase.} 
{Only the models accounting for the effects of the wind anisotropies retain enough angular momentum in their core to produce a Gamma Ray Burst (GRB). The chemical composition is such that a type Ic supernova event occurs.} 
{Wind anisotropies appear to be a key physical ingredient in the scenario leading to long GRBs.}
\keywords{Stars: evolution -- Stars: rotation -- Stars: abundances -- Stars: mass loss}
 \maketitle

%
%________________________________________________________________

\section{Introduction}

%Every day, one Gamma Ray Burst occurs in the whole observable Universe. They are among the most energetic events in the Universe after the Big Bang. 
%Knowing the progenitors of the GRB will give indications on how GRB might be used to probe the deep Universe. 
Gamma Ray Bursts (GRB) are extraordinary events which can be seen at very high redshifts. The record holder is currently the GRB050904 at z=6.29 (Cusumano et al.~\cite{Cu06}). The study of  the conditions for their appearance and the underlying physics may thus give important information on the early Universe.

In recent years many observations have indicated that long soft Gamma Ray Bursts (GRBs) are associated with the explosion of massive stars (see the review by Woosley \& Bloom~\cite{WB06}). In a few cases it has been possible to observe the type of the supernova  associated with the GRB event. In four well established cases (see the review by Della Valle~\cite{VA06}), the supernova was of type Ic (Galama et al.~\cite{GA98}; Mazzali et al.~\cite{MAZ06}). This means that the spectra do not show any hydrogen and helium lines. Such observations gave strong support to the collapsar model proposed by Woosley~(\cite{W93}). In this model, the progenitor of the GRB is a massive Wolf-Rayet (WR) star that collapses into a fast rotating black hole. Fast rotation is important since it allows part of the infalling matter to form a disk around the black hole. The disk acts as an efficient engine to extract gravitational energy. Part of this energy is used to power strong polar jets. The GRB arises from interactions in these jets. The short duration of the GRBs points toward a compact progenitor, {\it i.e.} a star having lost its outer envelope, a WR star possibly of the WO subtype (Woosley~\cite{W93};  Mazzali et al.~\cite{MAZ06}).

While this general scenario is now well accepted, the precise features of the long GRB progenitors are still unknown. First, not all WR stars will give birth to a GRB. Della Valle~(\cite{VA06}) for instance estimates that the number ratio of GRBs to type Ibc SNe is between 0.5-4\%. Hammer et al.~(\cite{Hetal06}) suggest that GRB progenitors are runaway massive stars ejected from compact massive star clusters.
Fruchter et al. (\cite{Fru06}) find that GRBs do not occur in similar environments to supernovae:
they found that the GRBs are far more concentrated in the very brightest regions of their hosts galaxies than are supernovae. Moreover, the GRB host galaxies are fainter and more irregular than the hosts of supernovae. These observations may point toward a very massive progenitor in metal poor environments. In seven cases the  metallicity of the galaxy hosting the GRB was measured or at least was given an upper limit. In all the cases, the metallicity was found inferior to one third solar (reported as a personal communication from Vreeswijk in Fruchter et al.~\cite{Fru06}). 

The progenitors of long GRBs represent a real challenge for stellar physics.
On one hand, it is necessary that the star loses a large fraction of its initial mass in order
to make a WO star. On the other hand, the star must 
 keep a high specific angular momentum $j$ ($> 10^{16}$ cm$^{-2}$ s$^{-1}$) in its  core for the collapsar model to work. These two constraints are relatively contradictory, because
 removing a lot of mass implies losing a lot of angular momentum. This is the difficulty 
 for models of GRB progenitors. 
%The loss of angular momentum is also
%influenced by the  instabilities triggered by rotation (shear instabilities, meridional circulation, %transport of the angular momentum by the magnetic fields).

Models accounting for the effects of rotation but without magnetic fields have no difficulty in preserving
enough angular momentum in their central regions (Heger et al.~\cite{HLW00}; Hirschi et al.~\cite{HMM05}). 
The  conditions of having  WO stars as progenitors is consistent with metal poor environments (see Hirschi et al.~\cite{HMM05}). However, these models may somewhat overestimate the frequency of GRBs and cannot reproduce
the observed rotation rate of young pulsars.
When  the magnetic fields are accounted for (Heger et al.~\cite{HWS05}) as proposed in the dynamo theory of Spruit~(\cite{Sp02}),  the formation of a collapsar is difficult, at least for not too extreme velocities. In that case, the magnetic field imposes solid body rotation during the Main Sequence phase, allowing efficient transfer and extraction of angular momentum from the core and the models have not enough angular momentum in their core to give birth to a collapsar.

%As indicated above, the formation of a collapsar requires at least three conditions to be fullfilled: 1) the central region must have retained a sufficiently high content of angular momentum; 2) The hydrogen and helium rich envelope must have been thrown away in the previous phases by stellar winds; 3) the core size at the pre-supernova stage must be massive enough for forming a black-hole.

A way to overcome this difficulty was proposed by Yoon \& Langer~(\cite{YL05}) and Woosley \& Heger~ (\cite{WH06}), who computed the evolution of very fast rotating massive stars with homogeneous evolution similar to the models computed by Maeder~(\cite{M87}). These stars remain in the blue part of the HR diagram and avoid the red supergiant phase. However, such models enter quite early into the WR phase due to the action of strong mixing and thus  have to escape the danger of losing too much angular momentum in the very strong WR stellar winds. To avoid this, a second crucial element has to  be accounted for:
the fact recently shown that the winds of WR stars are metallicity dependent (Crowther et al.~\cite{Cro02}; Vink \& Koter~\cite{VK05}), {\it i.e.} weaker at lower metallicity. When such mass loss recipes are used,  the homogeneous models may lead to collapsars.
Within this scenario, the rarity of the GRB events is due to the fact that the GRB progenitors only occur for  high initial stellar masses, for very high rotational velocities and for low metallicities.
%This does not prevent the stars to explode with modest masses (cf. Fig. \ref{mass}).

\begin{figure}
\resizebox{\hsize}{!}{\includegraphics[angle=00]{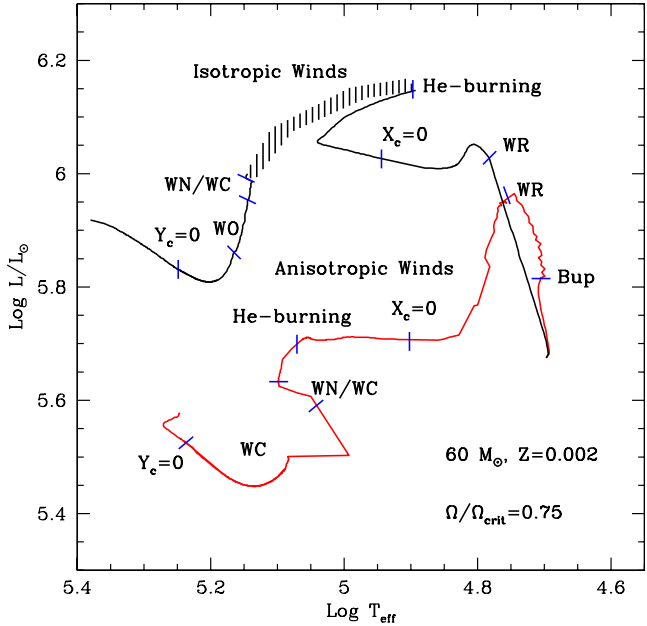}}
\caption{Evolutionary tracks for rotating models with and without wind anisotropies. The plotted effective temperatures correspond to the surface averaged effective temperature at the hydrostatic surface 
(no correction for the optical thickness of the wind). The beginning of various phases are indicated along the evolutionary sequences: {\bf Bup}, {\bf WR},
and $X_c=0$ indicate when the star encounters the critical or break-up limit, enters the WR phase and when core H-burning stops. The core He-burning phase begins at the point labeled {\bf He-burning}. The transition phase between the WNL and WC phases called the WN/WC phase corresponds to the portion of the tracks
comprised between the two tick marks with, in the middle, the label {\bf WN/WC}. After that point the star is a {\bf WC} star. The entrance into the {\bf WO} phase is indicated (if it occurs), as well as the endpoint of the core He-burning phase labeled by $Y_c=0$.}
\label{dhr}
\end{figure}

Models in the literature do not account for the effects of the wind anisotropies induced by rotation.
Here we present the first computation including magnetic field and wind anisotropies.
%some physical ingredients may be still lacking: 1) The prescriptions these authors use to account for the effects of rotation on mass loss implicitly suppose that the fast rotating stars have a uniform brightness, which is not the case. 2) They do not account for the effects of the wind anisotropies induced by rotation. 
%3)  The critical mass $M$ for vibrational instabilities of homogeneous models obeys the relation $\mu^2 M$    = constant (Ledoux 1941) where $M$ is the actual mass and $\mu$ the mean molecular weight. It has been shown by Maeder (A\&A, 147, 300, 1985) that when the ratio of hydrogen to helium at the surface becomes equal to about 0.3 the homogeneous star enters into the unstable regime and remains in this regime for the rest of its lifetime. These instabilities may trigger mass loss.
We revisit the above scenario (homogeneous evolution) accounting for this effect.
In Sect.~2, the model ingredients are presented. In Sect.~3 to 5 we discuss the effects of wind anisotropies on the evolution, rotation and chemical abundances. Conclusions are given in Sect.~6.

\section{The physical ingredients}

The stellar ingredients are the same as in Meynet \& Maeder~(\cite{MMXI}) with the
following exceptions: 1) For the mass loss rates we use the same prescriptions as in Yoon \& Langer (\cite{YL05}; their WR1 recipe during the WR phase).
%{\it i.e.} Kudritzki et al. (1989) and for the WR phase, their relation labeled WR1.
We suppose that the star enters the WR phase when the hydrogen surface abundance becomes
inferior to 0.3 in mass fraction; 2) The effects of the magnetic field are accounted for as in Maeder \& Meynet~(\cite{magn3}). Let us recall that in rotating models with magnetic field, the main mechanism
driving the transport of the chemical species is meridional circulation.

We calculate two 60 M$_\odot$  models with an initial composition $X=0.748$ and $Z=0.002$ with a value $\Omega/\Omega_{\rm crit}=0.75$ on the ZAMS
($\upsilon_{\rm ini}=523$ km s$^{-1}$). Note that such high initial velocities are required for at least two purposes:
first in order to have a well developed wind anisotropy, second such initial velocities seem to be required
to obtain a collapsar in rotating models with magnetic fields (see Yoon \& Langer~\cite{YL05}; Woosley \& Heger~\cite{WH06}). Faster rotation will likely give similar results to those shown here, while lower
initial rotation would probably show no important wind anisotropy effect and thus would not lead to a collapsar.
One model was computed assuming isotropic winds and one model with
anisotropic winds. The effects of the wind anisotropies are computed as in Maeder~(\cite{M02}).
The computations were followed until the end of the core He-burning phase. After that stage, the angular momentum of the core does not change much (see Hirschi et al.~\cite{HMM05}) and has already reached a value 
near the one it will have at the presupernova stage. This conclusion was based on models without a magnetic field and with a lower initial rotation. Woosley \& Heger~(\cite{WH06}), however show that the specific angular momentum at the Lagrangian mass of 3M$_\odot$ decreases by an order
of magnitude between the end of the core He-burning phase and the presupernova stage in models with a magnetic field and fast rotation. In similar models by Yoon and Langer~(\cite{YL05}), on the contrary, the decrease of the inner specific angular momentum between the end of the He-burning and the end of the core C-burning is very modest (cf. their Fig.~5). Thus, both our models (Hirschi et al.~\cite{HMM05}) and those of Yoon \& Langer~(\cite{YL05}) support a small change of the specific angular momentum in the advanced nuclear stages, in contrast with Woosley \& Heger~(\cite{WH06}). Theoretically, we do not see which mechanism (among those included) could efficiently remove the specific angular momentum from the core in the late nuclear stages, since the two most efficient processes ({\it i.e.} magnetic field and transport by meridian circulation) vanish or become inefficient in regions with $\mu$-gradients as strong as those near the core of massive stars.

As will be discussed below, the surface velocity of our ``wind anisotropic'' model reaches the critical limit.
At the critical limit, wind anisotropies are not accounted for and the loss of angular momentum corresponds to that the star
would have from a spherical wind. For a star of T$_{\rm eff} < 25000$ K, this would probably be wrong, because the opacity at the equator is higher due to the lower T$_{\rm eff}$ and
it would drive the matter outward in the equatorial plane. For very hot stars with T$_{\rm eff} > 50000$ K, the star keeps a strong polar wind at the critical limit, and some matter is ejected in the equatorial plane. It is difficult at the present
time to predict what would be the net effect on the angular momentum content of the star. However, at the critical limit,  it is in general sufficient that a small amount of mass is removed to bring the star away from the critical limit, {\it i.e.} within the regions of polar stellar winds. Therefore the combination of anisotropic winds and spherical winds at break-up may be an acceptable assumption, at least as long as the physics of break-up has not been more investigated.

%which reach the $\Omega\Gamma$-limit, it was pointed out (Maeder \& Meynet~\cite{MMVI}) that $\Gamma_\Omega(\theta)$ is essentially constant over the stellar surface and thus the above assumption may be appropriate. In addition, we notice that at critical velocity, it is in general sufficient that a small amount of mass is removed to bring the star away from the critical limit,
%{\it i.e.} within the regions of anisotropic stellar winds. Therefore the combination of anisotropic winds and spherical winds at break-up may be an acceptable assumption, at least as long as the physics of break-up has not been more investigated.}

%voir oversh.
 
\section{Effects of wind anisotropies on the evolution}

\begin{figure}
\resizebox{\hsize}{!}{\includegraphics[angle=00]{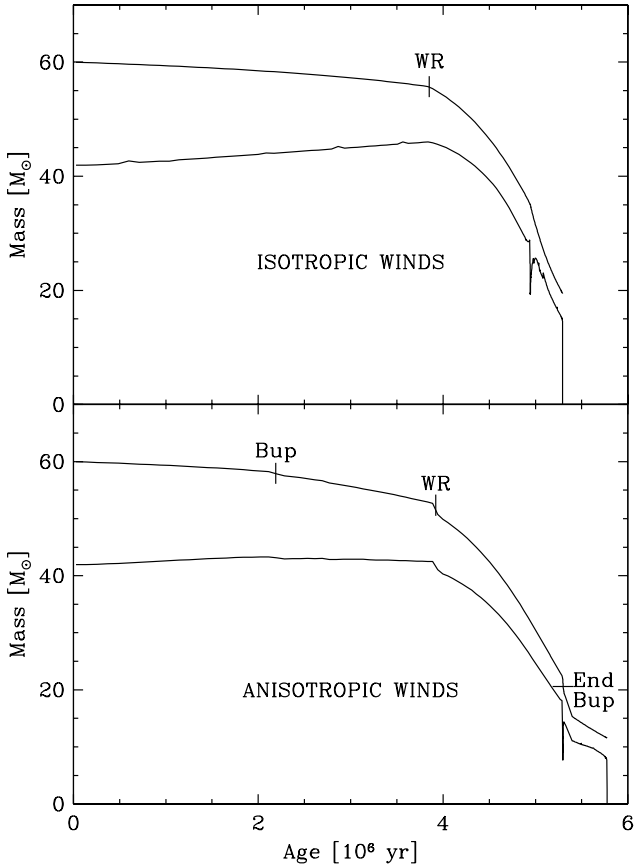}}
\caption{Evolution as a function of time of the total masses (upper curve) and the masses of the convective cores
(lower curve) for the models with isotropic winds (upper panel) and anisotropic winds (lower panel).
The isotropic model reaches $X_{\mathrm{c}}=0$ at $4.94 \times 10^6$ yr and  $Y_{\mathrm{c}}=0$ at $5.29 \times 10^6$ yr. For the anisotropic case, the corresponding ages are $5.29 \times 10^6$ yr and
$5.78 \times 10^6$ yr. The labels {\bf Bup}, {\bf WR} and {\bf End Bup} show the points when the star encounters the Break-up limit, enters the WR phase and evolves away from the Break-up limit.}
\label{mass}
\end{figure}

\begin{figure}
\resizebox{\hsize}{!}{\includegraphics[angle=00]{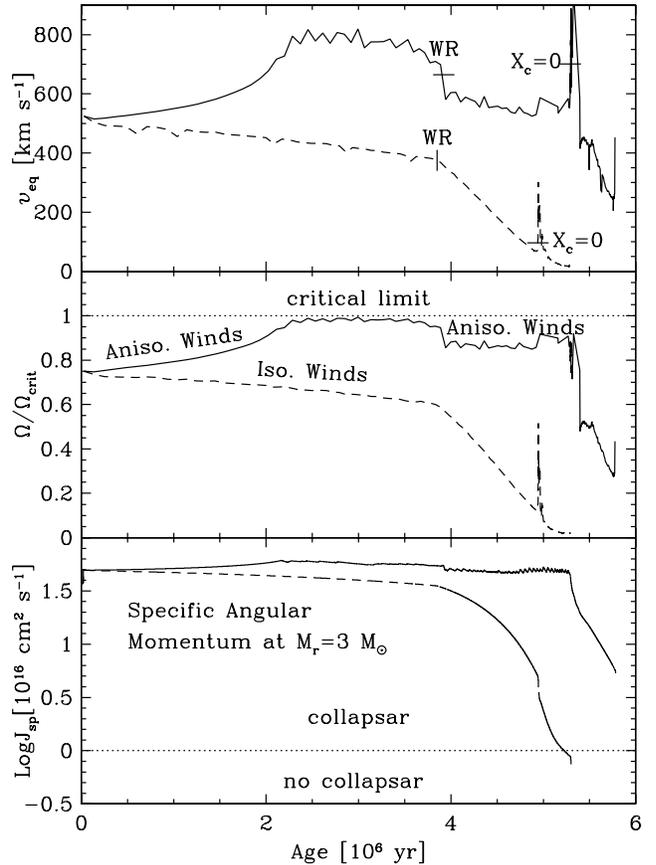}}
\caption{Evolution as a function of time of the equatorial velocity (top) and of the ratio $\Omega/\Omega_{\rm crit}$ at the surface (middle). The bottom panel shows the evolution of the specific angular momentum at the Lagrangian mass coordinate $M_r=3$ M$_\odot$.}
\label{ne}
\end{figure}

\begin{figure}
\resizebox{\hsize}{!}{\includegraphics[angle=00]{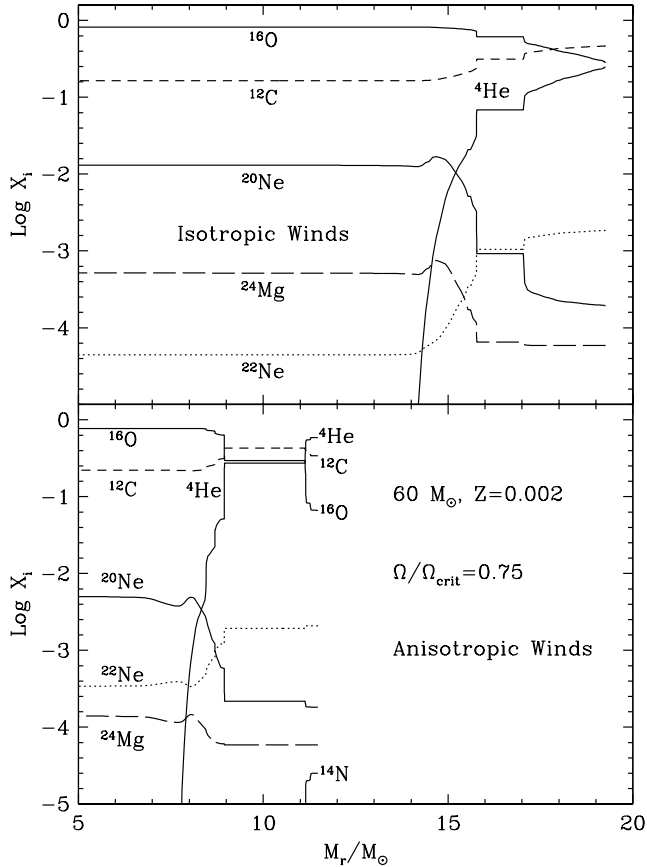}}
\caption{Abundances (in mass fractions) as a function of the Lagrangian mass of various
 elements inside  stellar models without (top) and with (bottom) wind anisotropies when $Y_{\mathrm{c}}=0$.}
\label{ab}
\end{figure}

Since homogeneous evolution may have observational counterparts (see e.g. Walborn et al.~\cite{Wa04}), we discuss some caracteristics of our evolutionary tracks in the HR diagram (see Fig.~\ref{dhr}).
Due to their fast rotation and thus strong internal mixing, the
stars remain in the blue part of the HR diagram. Both models follow qualitatively the same evolution, going from an O-type star phase through a WR phase, although with 
some striking differences:

%1) At the beginning of the MS phase, the models with wind anisotropies slightly evolve to the red with respect to the model without the wind anisotropies\footnote{Let us recall here that we have plotted here surface averaged effective temperatures}. This results from the fact that when anisotropic winds are accounted for, less angular momentum is removed from the star at the surface. Thus the outer layers are more deformed implying a lowering of the averaged effective temperature.

1) The ``anisotropic'' model encounters the break-up or critical limit during the MS phase. This is of course a natural consequence of the smaller amount of angular momentum lost. 
%\footnote{Note that at the critical limit, we remove the wind anisotropies and adopt
%the isotropic mass loss rate in order to allow the star to evolve away from the critical limit. Away from the %critical limit, when $\Omega/\Omega_{\rm crit}=0.95$, the anisotropic wind is activated again until the %critical limit is reached again.}.

2) The entrance into the WR phase occurs  at about the same evolutionary stage in both models, {\it i.e.} with a mass fraction of hydrogen at the center $X_c=0.28$ and $X_c=0.29$ in the ``isotropic''
respectively ``anisotropic'' wind stellar models and with an actual mass of 52.4 M$_\odot$, respectively 55.7 M$_\odot$. However in the case of the anisotropic model, the first part of the WR phase (between the labels WR and $X_c=0$ in Fig.~\ref{dhr}) shows a steep decrease in luminosity, while in the isotropic model the corresponding change is limited. The reason is that in the anisotropic model during this phase, the star continues to be at the critical limit (see Fig.~\ref{ne}), where it loses  more mass than the isotropic
 model which is far from this limit. Thus, due to the mass-luminosity relation followed by homogeneous models, the ``anisotropic'' model undergoes a more pronounced luminosity decrease. When $X_c=0$, the isotropic model has an actual mass of 35.1 M$_\odot$, while the anisotropic model has only 22.1 M$_\odot$ (see also Fig.~\ref{mass}).

3) The transition between the end of the core H-burning and the beginning of the core He-burning phase
is different in the two models (cf. Fig. \ref{dhr}). This comes from the fact that the isotropic model still has a small H-rich envelope (mass fraction of hydrogen at the surface $X_s=0.07$), while the anisotropic model is nearly a pure
He-core ($X_s=0.001$).  When, at the end of the core H-burning phase, a small H-rich envelope is present, 
a phase of expansion follows the contraction of the star. The expansion phase is
stopped when the star encounters  the $\Omega\Gamma$-limit. This corresponds to the point labeled ``He-burning'' along the isotropic track in Fig.~\ref{dhr}. This produces an enhancement of the mass loss for a short time. After this phase, the star describes some back and forth movements in the HR diagram, which we have schematically represented by a hatched zone in Fig.~\ref{dhr}. For the anisotropic model, the core simply contracts. More mass is lost
since  contraction bring more mass above the critical limit. This increase in the mass loss finally makes
the star evolve definitively away from the critical limit. 

4) In both models, the end of the core He-burning phase occurs away from the critical limit
(Fig. \ref{ne}). The evolutions of the two models (after the label WN/WC) differ mainly in the luminosity: the anisotropic model has a luminosity about 0.30 dex below that of the isotropic model, 
because the first model has lost much more mass than the second one. The final masses of the isotropic and anisotropic models are 19.3 and 11.5 M$_\odot$ respectively. 

\section{Evolution of rotation}
%manque dessin M vs age ou table
Figure \ref{ne} shows the evolution  as a function of time of the equatorial velocity at the surface, of the ratio of the surface angular velocity to the critical velocity and of the specific angular momentum at the Lagrangian mass coordinate $M_r=$ 3 M$_\odot$.
 In the anisotropic model, the surface velocity increases first and then remains at the critical 
 limit during the rest of the core H-burning phase. When the star enters the WR phase, the star continues to encounter the critical limit, however the higher mass loss rates maintain the star on average at a slightly greater distance from the critical limit. In the isotropic model, the surface velocity decreases and approaches the critical limit only during a very short episode in the transition phase between the core H- and He-burning phase. In the bottom panel of Fig.~\ref{ne}, one  sees the importance of the wind anisotropies on the specific angular momentum at $M_r= 3$ M$_{\odot}$. 
 As long as the stellar models do not lose too much mass by stellar winds, {\it i.e.} before the WR phase, the specific angular momentum at $M_r= 3$ M$_\odot$ remains nearly constant.% with even a slight increase in the
%case of the anisotropic model.
 When the star enters the WR phase, the specific angular momentum strongly decreases in the isotropic model, while, when anisotropic winds are considered, the specific angular momentum shows nearly no decrease. This allows the star to enter the core He-burning phase with
a high angular momentum content. At $M_r=3$  M$_\odot$, the specific angular momentum is about 9 times larger in the anisotropic model than in the isotropic one. Moreover, since the anisotropic model has lost more mass (but much less angular momentum),   
the mass loss rates  in the core He-burning phase (due to their specific dependence on   mass)
are lower than for the isotropic model. This allows the anisotropic model to end its evolution with sufficient angular momentum to give birth to a collapsar.
On the contrary the isotropic model has too low a central specific angular momentum to be a progenitor of a long GRB. 

\section{Chemical abundances}   

In Fig.~\ref{ab}, the variations of the abundances in the two 60 M$_\odot$  models are shown when 
Y$_{\mathrm{c}}=0$. The isotropic model has less He than CO at the surface, while for the anisotropic model
it is the opposite. This at first sight might be surprising. Indeed, why is the model losing the greatest quantity of mass, ending with more helium at the surface? The reason is  
that  when the star loses more mass, it enters the WC phase at an earlier stage of the core He-burning phase,
{\it{i.e.}}
when smaller amounts of helium have been processed into carbon and oxygen (Smith \& Maeder~\cite{SM91}). This is what happens here.
The anisotropic model enters the WC phase when the mass fraction of helium at the center $Y_c=0.93$, while the isotropic model enters the WC phase with $Y_c=0.57$. Taken at face value, the fact that the anisotropic model produce a WC progenitor differs with the type Ic supernova associated with long GRB. However if we integrate the total quantity of helium  ejected (supposing that all the He-rich layers are ejected at the time of the GRB event),  we obtain not too different values, about 0.53 M$_\odot$ for the isotropic model and 0.77 M$_\odot$ for the anisotropic one. Such quantities of helium, compared to current supernovae,  may escape detection during a type Ic event and agree with the idea that type Ic supernovae either eject He-deficient material or fail to non thermally excite their helium
(Woosley \& Eastman~\cite{WE97}). 

%\section{Conclusions}

\section{Conclusions}

We have shown that wind anisotropies play an important role in the scenario leading to collapsars and then to long GRBs. Wind anisotropies help maintain a high content of angular momentum in the stellar core, while 
they make  the star lose very large amounts of mass. The anisotropic model loses more mass than the isotropic model, because the anisotropic model remains near the critical limit for most of the core H-burning phase. Interestingly, despite these important losses of matter, the anisotropic model ends its stellar lifetime
with high surface velocities, between 200-400 km s$^{-1}$, while the isotropic model shows very modest values (a few tenths of km s$^{-1}$, see Fig.~\ref{ne}). At very low metallicity, GRB progenitors are expected to show  high surface ratios of [Ne/Fe]. 
Usually
the abundance of $^{22}$Ne observed at the surface of a WC or WO star is equal to the initial
CNO abundance. In the case of primary nitrogen production (expected to be abundantly produced
in very metal poor fast rotating stars), the abundance of $^{22}$Ne might be much higher that the initial
CNO abundance. Thus Wolf-Rayet
progenitors of collapsars at very low metallicity should show, together with a high surface rotational
velocity, a very high [Ne/Fe] ratio.
%It has to be stressed that such models would predict non isotropic distribution of mass in the
%GRB neighborhood. 
In a further work, we will explore the consequences of wind anisotropies on a larger ranges of masses, metallicities and initial rotation velocities. 
%We shall also show how the matter lost in previous stellar winds episodes is distributed at the time of the GRB event.    
%critical equatorial?
%why no long at the critical limit.

%\end{acknowledgements}

\end{document}